# Gamma-ray Observatory INTEGRAL reloaded

Edward P.J. van den Heuvel

Anton Pannekoek Institute of Astronomy, University of Amsterdam

**A new lease on life was given to ESA's International Gamma-ray Astrophysics Laboratory because of its unique capability to identify electromagnetic counterparts to sources of gravitational waves and ultra-high energy neutrinos.**

Sometimes it happens in astronomy that an instrument that was designed for a certain purpose, turns out to be ideally suited for a completely different task, not foreseen when the instrument was designed. This is now the case with ESA's International Gamma-Ray Laboratory INTEGRAL, which was designed for the discovery, precise location and high-resolution γ-ray spectroscopy of cosmic high-energy sources in the energy range 15 keV to 10 MeV, with concurrent source monitoring in the X-ray and optical energy bands. After more than 14 years in space, all its 4 instruments are still functioning very well. In the last years, the community of scientists working with the satellite realized that INTEGRAL is perfectly suited for the follow-up and location of Gravitational Wave (GW) sources and ultra-high-energy (UHE) neutrino sources in the Universe. The detection of a counterpart at electromagnetic wavelengths will be very important for providing a precise position, enabling redshift determination, association to a host galaxy, and important astrophysical context. With Advanced LIGO, VIRGO and other GW antennas coming on line, the coming years are expected to be the "Golden Age" of GW astronomy.

Earth's atmosphere shields us from γ-rays from the Universe. For this reason, γ-rays from cosmic sources can only be observed with satellites. Participants in INTEGRAL are all member states of ESA, plus the Russian Federation and United States. The only rocket capable to lift the 4 tons satellite into its wide 3-day eccentric orbit above the radiation belts, was Russia's largest rocket, the Proton, which provided a perfect launch on October 17, 2002.

The four scientific instruments of INTEGRAL's payload module weigh 2 tons, making this payload the heaviest ever placed in a wide orbit by ESA. The two main instruments, which account for 90% of this weight, both have large ~30x30 degrees fields of View (FOV), and look in the same direction on the sky. They are the high-resolution spectrometer SPI ("Spectrometer for INTEGRAL"; energy resolution E/ΔE= 500 in the range 18keV to 10 MeV), and the imaging telescope IBIS ("Imager on-Board the INTEGRAL Satellite"), which gives the sharpest γ-ray images yet seen from astronomical targets (FWHM = 12 arc minutes), locating objects with a precision of down to 30 arc seconds. The SPI spectrometer detects γ-ray spectral lines emitted by decaying radioactive nuclei. Figure 1 shows as an example the detected γ-ray lines at 847 and 1238 keV, due to radioactive cobalt ($^{56}$Co) in Supernova 2014J, the first time ever this decay was observed (ref. 1). Two other instruments, an X-ray monitor and an optical camera, also looking in the same direction, help to identify the γ-ray sources and pursue complementary science. SPI and IBIS are equipped with active Anti-coincidence shields (ACSs) to reduce cosmic-ray induced background. These massive shields (512 and 115 kg of Bismuth-Germanate for SPI and IBIS, respectively) are ideal omni-directional detectors of cosmic flashes of γ-rays, and therefore: for the location of GW and neutrino sources. They detect on average some 200 Gamma-Ray Bursts (GRB) per year.

**BOX 1: Some of INTEGRAL's breakthrough contributions of the past 15 years are:**

- Discovery of over 600 new hard-X and soft γ-ray sources, more than doubling the known number of such sources (ref.2); identification and characterization of over 250 of the new sources; among these: two new classes of high-mass X-ray binaries (HMXBs): the highly-obscured systems, and the Supergiant Fast X-ray Transients; these discoveries quadrupled the known number of HMXBs with blue supergiant donor stars.
- First large-scale sky map at 511 keV, produced by positron annihilation, showing the presence of large amounts of anti-matter in the central parts of our Galaxy (ref.3);
- First-ever detection of $^{44}$Ti decay lines from the core-collapse supernova SN1987A (ref.4).
- Discovery of (variable) polarization of the γ-ray emission of the Crab Nebula (ref. 5), and of γ-ray polarization of the black-hole source Cygnus X-1 and of Gamma-Ray Bursts (GRBs).
- Proof of the galaxy-wide origin of $^{26}$Al, from which the current Galaxy-wide rate of core-collapse supernovae could be derived (ref. 6); insight in feedback from massive stars from the Doppler-kinematics of the $^{26}$Al decay line, and in supernova physics, from the observed $^{26}$Al/$^{60}$Fe ratio.
- First measurement of the fraction of Compton-thick Active Galactic Nuclei (AGN) in a complete AGN sample (ref. 7), and discovery of record-distance QSOs in hard X-rays.  **{end of box}**

Prior to September 2015, the only certain sources of strong bursts of gravitational waves observable with Advanced LIGO, were expected to be the mergers of close double neutron stars. We know at least 16 of such systems in our Galaxy, about half of which will merge on a timescale of a few hundred million years. On this basis and assuming similar systems to be present in all galaxies, as much as ~ 40 GW detections from merging double neutron stars may be observable per year in the period 2017-2018, out to ~ 120 Mpc distance (ref. 8). Merging double black holes will produce stronger GW signals, that can be observed to much larger distances, but until September 2015 the existence of close double black holes was not known with certainty. Since we now know that close double black holes do exist in the Universe (ref. 9), we can also be quite certain that close binaries consisting of a neutron star (NS) and a black hole (BH) do exist, and their GW detection rate might be higher than for double NSs (ref. 8). The mergers of double neutron stars are expected to produce a strong burst of γ-rays, known as a short Gamma-Ray Burst (GRB), with a duration of less than a few seconds, and the same is expected for NS-BH mergers. Whether also double black hole mergers might produce a γ-ray signal is uncertain, although models for this are conceivable (e.g.: ref. 10).

The sources of UHE cosmic neutrinos are also expected to be γ-ray sources, as in these sources particles are accelerated to highly relativistic energies, a process that also leads to γ-ray emission. Possible suggested sources are: supernova explosions and the relativistic jets of BL Lac objects (AGNs) and of objects producing GRBs, such as "collapsars" (stellar black holes surrounded by a short-lived accretion disk), and merging of double neutron stars. The Ice Cube detector records of order 15 astrophysical UHE neutrino events per year, of which one quarter are track detections, with a positional accuracy of order ± 1°, and three quarters are shower events with a positional accuracy of order ± 15° (ref. 14 and figure 2).

The unique large effective area of the ACSs of SPI and IBIS, make them above 75 keV most sensitive and virtually omni-directional (outside the telescopes' field of view) γ-ray detectors, providing count rates with 50 ms time resolution. Thanks to its long orbit (presently 2.7-days), the entire sky is accessible for continuous observations of up to 2 days. If the flash coincides in time with a GW or neutrino event, the satellite can be turned to the location on the sky where the GW-antenna or neutrino-telescope has roughly located the source (see figure 2). To improve the chances for finding counterparts – or placing stringent upper limits – INTEGRAL has for the coming years prioritized time-domain astronomy. The Target of Opportunity strategy of INTEGRAL provides rapid follow-up observations (response times from a few hours to about one day) after alerts from the GW or neutrino networks. The large FOVs of the two γ-ray instruments cover the error boxes of the LIGO-VIRGO network, i.e. 100 deg$^2$ (to be reached in 2017), as well as those of the neutrino observatories (fig.2), such that potentially the sources can be located and identified. In addition, during the coming years one expects several GW events (including double black hole mergers) to take place directly within the IBIS and SPI FOVs, such that an immediate source identification may ensue.

Realizing this potential INTEGRAL, like other major astronomical observatories, has teamed up with the LIGO-VIRGO collaboration and the IceCube collaboration, to look for possible counterparts of GW and UHE neutrino events. INTEGRAL has set the first stringent upper limit ($E_{gamma}/E_{GW} < 10^{-6}$, implying $L_\gamma < 7\times10^{46}$ ergs/s, ref.11) on direct γ-rays coincident with the double black hole merger event GW150914 (see figure 2), and on γ-rays from UHE neutrino events (e.g. ref. 12).

These might well be the first steps towards finding electromagnetic counterparts of these non-electromagnetic cosmic messengers.

**Acknowledgements**: I thank Erik Kuulkers and Peter Kretschmar for comments on an early draft of this paper and Erik Kuulkers also for providing figure 2 (right) and other important information.

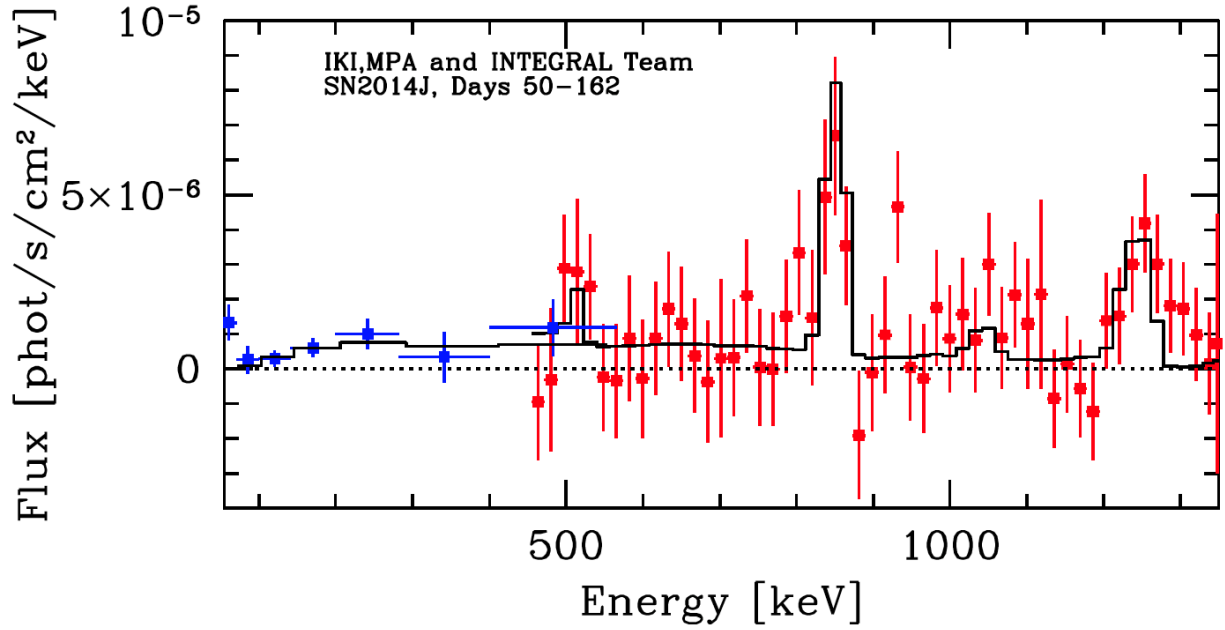

Figure 1: Combined IBIS/SPI spectrum for the 'late' period (50-162 days after the explosion) of Supernova 2014J. One of the best-fitting models is shown by the black line. The most prominent signature of the $^{56}$Co decay are the lines at 847 and 1238 keV. [Figure from ref. 1].

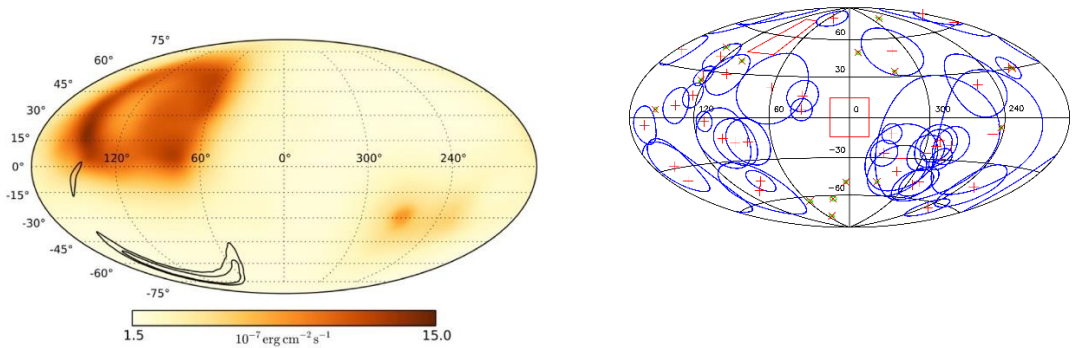

Figure 2: Left: The most accurate GW150914 event localization drawn as black contour regions at 50% and 90% confidence in Galactic coordinates (ref. 9 and 13). The SPI-ACS flux upper limit in 1 sec for a characteristic short hard GRB spectrum over the entire sky is also shown (darkest parts are around the observing direction of SPI). [Figure from ref. 11]. Right: arrival directions of 54 UHE neutrinos observed in 4 years (2010-2014) by the IceCube detector, in Galactic coordinates (figure constructed by Dr. E. Kuulkers on the basis of the data in ref: 14). Shower-like events are marked with + with corresponding error regions in blue, those containing tracks with x and green, respectively. The FOV of IBIS is shown in red in the middle and top left.